\def\Statusstring{}

\documentclass[12pt,letter]{article}

\usepackage{amsmath, amsthm, amssymb}
\usepackage{amssymb,epsfig}

\makeatletter

\gdef\@punct{.\ \ }  
\def\@sect#1#2#3#4#5#6[#7]#8{%
  \ifnum #2>\c@secnumdepth
     \def\@svsec{}
  \else
     \refstepcounter{#1}\edef\@svsec{%
     \ifnum #2>0{{\csname the#1\endcsname}}.\fi%
    \hskip .5em}
  \fi
  \@tempskipa #5\relax
  \ifdim \@tempskipa>\z@
     \begingroup #6\relax
       \@hangfrom{\hskip #3\relax\@svsec}{\interlinepenalty \@M #8\par}
     \endgroup
     \csname #1mark\endcsname{#7}
     \addcontentsline{toc}{#1}{\ifnum #2>\c@secnumdepth\else
          \protect\numberline{\csname the#1\endcsname}\fi#7}
  \else
     \def\@svsechd{#6\hskip #3\@svsec #8\@punct\csname
#1mark\endcsname{#7}
     \addcontentsline{toc}{#1}{\ifnum #2>\c@secnumdepth \else
          \protect\numberline{\csname the#1\endcsname}\fi#7}}
  \fi
  \@xsect{#5}}

\def\@ssect#1#2#3#4#5{\@tempskipa #3\relax
  \ifdim \@tempskipa>\z@
     \begingroup #4\@hangfrom{\hskip #1}{\interlinepenalty \@M
#5\par}\endgroup
  \else \def\@svsechd{#4\hskip #1\relax #5\@punct}\fi
  \@xsect{#3}}

\def\qed{\hskip 3pt \hbox{\vrule width4pt depth2pt height6pt}}

\newtheorem{Lemma}{Lemma}

\newtheorem{Theorem}[Lemma]{Theorem}
\newtheorem{Proposition}[Lemma]{Proposition}
\newtheorem{Corollary}[Lemma]{Corollary}

\begin{document}

\title{The regular number of a graph \thanks{This work was presented in part at the Annual Conference of ADMA (Academy of Discrete Mathematics and Applications), NIT Calicut, India, June 2011.} }
\author{Ashwin Ganesan and Radha R. Iyer \\ \textnormal{\small Department of Mathematics}
\\ \textnormal{\small Amrita School of Engineering, Amrita Vishwa Vidyapeetham,}
\\ \textnormal{\small Amritanagar, Coimbatore 641112, India.}
\\ \textnormal{\tt {\small \{ashwin.ganesan@gmail.com, g\_ashwin@cb.amrita.edu\},{ r\_radhaiyer@cb.amrita.edu}}}
}

\date{}

\maketitle

\vspace{-7.5cm}
\begin{flushright}
  \texttt{\Statusstring}\\[1cm]
\end{flushright}
\vspace{+5.5cm}

\begin{abstract}
\noindent  Let $G$ be a simple undirected graph.  The regular number of $G$ is defined to be the minimum number of subsets into which the edge set of $G$ can be partitioned so that the subgraph induced by each subset is regular. In this work, we obtain the regular number of some families of graphs and discuss some general bounds on this parameter.
Also, some of the lower or upper bounds proved in \cite{Kulli:Janakiram:Iyer:2001} are shown here to hold with equality.
\end{abstract}




\section{Introduction}

Let $G=(V,E)$ be a simple, undirected graph.  The \emph{regular number} of a graph $G$, denoted by $r(G)$, is
the minimum number of subsets into which the edge set of $G$ can be partitioned so that the subgraph induced by each subset is regular.  This quantity was introduced in \cite{Kulli:Janakiram:Iyer:2001}. Nonempty subsets $E_1,\ldots,E_r$ of $E$ are said to form a \emph{regular partition} of $G$ if the subgraph induced by each subset is regular.  If the number of these nonempty subsets is minimal, we say that these subsets form a \emph{minimum regular partition} of $G$.  For basic notation and terminology on graph theory used in this paper, we refer to \cite{Bollobas:2002}.

Note that in the definition of $r(G)$, we do not require that each element of the minimum regular partition of $G$ span the entire graph.  It can certainly be the case that a particular subset of $E$ in the minimum regular partition of $G$ does not cover all the vertices.  We assume that the regular number of an empty graph is 0 since it contains no edges.  Note that the graph on six vertices consisting of the 5-cycle plus one isolated vertex is irregular but has regular number equal to 1.  In the sequel, we assume throughout that $G$ is a simple, undirected graph, without loops or parallel edges and containing at least one edge.

The chromatic index (also known as the edge coloring number) of a graph is the minimum number of subsets into which the edge set of $G$ can be partitioned so that each subset is a matching.  Thus, while the regular number of a graph is the minimum size of a partition of the edge set into regular subgraphs, the chromatic index of a graph is the minimum size of a partition of the edge set into 1-regular subgraphs.  Hence, $r(G)$ is bounded from above by the chromatic index of $G$.  While adding edges to a graph can never reduce its chromatic index, adding edges to a graph can sometimes reduce its regular number.  For example, $r(K_4 -e) = 2$, whereas $r(K_4)=1$.  

This paper is organized as follows.  In Section~\ref{sec:examples}, we compute the regular number of some families of graphs.  We extend the results given in \cite{Kulli:Janakiram:Iyer:2001}; some of the bounds proved there are shown here to hold with equality.  In Section~\ref{sec:bounds}, we provide some general bounds for the regular number.  Section~\ref{sec:conclusions} contains some concluding remarks.  This paper also serves to correct some of the errors and complete the proofs in \cite{Kulli:Janakiram:Iyer:2001} (such as Proposition 2, 3 and 4 there), where some results are stated with equality but the proofs show only an upper or lower bound.

\section{Regular number of some families of graphs}
\label{sec:examples}

In this section we compute the regular number of some families of graphs, such as the wheel graphs, trees, and complete bipartite graphs.

\begin{Proposition} For $p \ge 5$, let $W_p$ be the wheel graph on $p$ vertices.  Then, $r(W_p) = \lceil{\frac{p}{2}} \rceil$.
\end{Proposition}

\noindent \emph{Proof:}
Let $v_1,v_2,\ldots,v_p$ be the vertices of $W_p$, where $v_p$ is the center vertex of degree $p-1$.  Let $e_i$ be the edge $\{v_i,v_{i+1}\}$, $i=1,\ldots,p-2$, and let $e_i'=\{v_i,v_p\}$, $i=1,\ldots,p-1$, and $e_{p-1} = \{v_{p-1},v_1\}$.  To determine $r(W_p)$, we consider two cases, depending on whether $p$ is odd or even:

(i) Suppose p is odd.
Then the subsets $E_1 = \{e_1,e_1',e_2'\}$, $E_2=\{e_3,e_3',e_4'\}$, $E_3=\{e_5,e_5',e_6'\}$, $\ldots$,$E_{k-1} = \{e_{p-2},e_{p-2}',e_{p-1}'\}$, and $E_k=\{e_2,e_4,e_6,\ldots,e_{p-1}\}$ form a regular partition of $W_p$, where $k = \frac{p+1}{2} = \left \lceil \frac{p}{2} \right \rceil$. Hence, $r(W_p) \le \left \lceil \frac{p}{2} \right \rceil$.  We now prove the opposite inequality.  Let $E_1,\ldots, E_r$ be a regular partition of $W_p$.   The subgraph $\langle E_i \rangle$ induced by $E_i$ cannot be 3-regular since that would require that $\langle E_i \rangle = W_p$, a contradiction since $ \langle E_i \rangle$ is required to be regular whereas $W_p$ is irregular.  Hence, each $ \langle E_i \rangle$ is either 2-regular or 1-regular.  Thus, the $p-1$ edges incident to $v_p$ must be a part of at least $\frac{p-1}{2}$ different subsets $E_i$, implying that $r(W_p) \ge \frac{p-1}{2}$.  Furthermore, if any subgraph $ \langle E_i \rangle$ is 2-regular, say containing vertex $v_2 \ne v_p$, then a third edge incident to $v_2$ must be a part of some other 1-regular $ \langle E_j \rangle$, where $j \ne i$.  Thus, at least one of the $ \langle E_i \rangle$'s has to be 1-regular.  Since, the degree of $v_p$ is $p-1$, if at least one of the $\langle E_i \rangle$'s is 1-regular, we require $r(G) \ge p/2$.  Hence, $r(G) \ge \left \lceil \frac{p}{2} \right \rceil$.

(ii) Suppose p is even.
Let $E_1,\ldots,E_k$ be a regular partition of $G$.  As before, each $\langle E_i \rangle$ is either 1-regular or 2-regular.  Hence, since the degree of $v_p$ is $p-1$, we have that $r(G) \ge \frac{p-1}{2}$.  Since $p-1$ is odd, this implies that $r(G) \ge p/2$.  To prove the opposite inequality, we provide an explicit description of the subsets $E_i$ that form a regular partition of $G$ into $k=p/2$ subsets. Let $E_1,E_2,\ldots,E_{k-2}$ be the 3-cycles $\{e_1,e_1',e_2'\}$, $\{e_3,e_3',e_4'\}$, $\{e_5,e_5',e_6'\}$, $\ldots$,$\{e_{p-5},e_{p-5}',e_{p-4}'\}$, let $E_{k-1}$ be the 4-cycle $\{e_{p-3},e_{p-2},e_{p-1}',e_{p-3}'\}$, and let $E_k$ be the perfect matching $\{e_2,e_4,e_6,\ldots,e_{p-4},e_{p-2}',e_{p-1} \}$.  This partition is regular.
\hfill\qed

\bigskip
\begin{Proposition}
Let $T$ be any tree.  Then $r(T) = \Delta(T)$, where $\Delta(T)$ denotes the maximum degree of $T$.
\end{Proposition}

\noindent \emph{Proof: }
Let $E_1,\ldots,E_k$ be a regular partition of $T$.  Then each $\langle E_i \rangle$ is necessarily 1-regular, for if any $\langle E_i \rangle$was s-regular for some $s \ge 2$, then that would imply that $\langle E_i \rangle$ has cycles and therefore $T$ also has cycles, a contradiction.  Thus, each $E_i$ is a matching of $T$.  Hence, $r(T)$ is the chromatic index (also known as the edge coloring number) of $G$.  Since $T$ is bipartite, it follows from a well-known result \cite{Bollobas:2002} that its chromatic index is equal to its maximum degree.
\hfill\qed

\bigskip
\begin{Proposition}
Let $K_{m,n}$ be the complete bipartite graph, with $m \le n$.  Then,
$$ \left \lceil \frac{n}{m} \right \rceil \le r(K_{m,n}) \le \left\lfloor {\frac{n}{m}} \right\rfloor +m. $$
If $m$ divides $n$, then $r(K_{m,n})=n/m$.
\end{Proposition}

\noindent \emph{Proof: }
First note that $r(K_{1,t})=t$.  Then, in any regular partition $E_1,\ldots,E_k$ of $K_{m,n}$, $|E_i|^2 \le m^2$.  Hence, $k \ge |E|/m^2 = n/m$, which establishes the lower bound.  Next, consider the case where $m$ divides $n$. By setting each $E_i$ to be the $m^2$ edges of a $K_{m,m}$ in a manner that the $E_i$'s are disjoint, we get a regular partition of $K_{m,n}$ into $n/m$ subsets.  Hence, $r(K_{m,n})=n/m$ when $m$ divides $n$. Now suppose $n=mk+d$, where $d \in \{1,\ldots,m-1\}$.  Then, using the preceding procedure, we first construct a regular partition of $K_{m,mk}$ into $k$ subsets.  This still does not include the edges of one $K_{m,d}$, for which we still need to construct a regular partition.  Thus,
$$r(K_{m,n}) \le k + r(K_{d,m}).$$  Since the regular number of a graph is upper bounded by its chromatic index, and since the chromatic index of a bipartite graph equals its maximum degree, we get $r(K_{d,m}) \le m$.
\hfill\qed


\bigskip \noindent Let $m \le n$ and $n=mk+d$, where $d \in \{1,\ldots,m-1\}$.
It seems to us that $$r(K_{m,n}) = \left \lfloor \frac{n}{m} \right\rfloor + r(K_{d,m}),$$ i.e. equality might hold in the upper bound in the Lemma above.  In particular, this implies $r(K_{n-1,n}) = n$.  If this is true, then there is a quick, recursive procedure for determining the exact value of $r(K_{m,n})$.  The recurrence relation arising from such a result has the initial condition $r(K_{1,d})=d$ and $r(K_{d,d}) = 1$.  For example,
$$r(K_{10,74}) = 7 + r(K_{4,10}) = 7+2+r(K_{2,4}) = 11.$$
An interesting problem is to design an algorithm to decompose a given bipartite graph into a minimum number of regular subgraphs.

\bigskip \noindent We state the inequality discussed in the previous proof:

\begin{Corollary}
Suppose $m \le n$ and $n=mk+d$, where $d \in \{1,\ldots,m-1\}$.  Then,
$$r(K_{m,n}) \le k + r(K_{d,m}).$$
\end{Corollary}

\bigskip \noindent Preliminary calculations seem to indicate that this bound holds with equality.  We prove this next for the special case where $m=3$.

\begin{Theorem}
For the complete bipartite graph $K_{3,n}$, where $n \ge 1$, we have that $r(K_{3,n}) = n/3$ if $n \equiv 0$ ~(mod~3), and $r(K_{3,n}) = \left \lfloor \frac{n}{3} \right \rfloor + 3$ if $n \equiv 1,2$~(mod~3).
\end{Theorem}

\noindent \emph{Proof: }
We make a counting argument that relates the number of edges in the graph to the degree sequences of certain subgraphs.  We proceed by induction on $n$.  The proof has three parts.  First we settle the case $n \equiv 0~$(mod~3).  Then, we prove the basis step of the induction for $n \le 5$.  Finally we finish with the inductive step.  Let $G=K_{3,n}$ have bipartition $(X,Y)$, where $X=\{x_1,x_2,x_3\}$ and $Y=\{y_1,\ldots,y_n\}$.

First assume $n \equiv 0$~(mod~3).  Clearly, $r(K_{3,3})=1$.  Fix $n$ to be a multiple of 3 and assume the result holds for smaller values of $n$.  In any regular partition $E=\bigcup E_i$, we have that $|E_i| \le 9$.  Hence, $r(G) \ge |E|/9 = n/3$.  To prove the opposite inequality, let $E_1$ be the 9 edges of the $K_{3,3}$ subgraph induced by $X$ and $\{y_{n-2},y_{n-1},y_n\}$.  Then, $r(K_{3,n}) \le 1+r(K_{3,n-3})$ = $1+ (n-3)/3 = n/3$.  For the rest of the proof, we now assume that $n \equiv 1,2$~(mod~3).

We now prove the assertion for $n \le 5$.  It is clear that $r(K_{3,1})=3$.  To prove that $r(K_{3,2})=3$, assume, by way of contradiction, that $E=E_1 \cup E_2$ is a regular partition of $K_{3,2}$.  If $\langle E_i \rangle$, the subgraph induced by $E_i$, is 1-regular, then $|E_i| \in \{1, 2, 3\}$, and if $\langle E_i \rangle$ is 2-regular, then $|E_i| \in \{4\}$.  Furthermore, if $E_i$ is 1-regular, it can have at most 2 independent edges.  Hence, $|E_1|+|E_2|=6$ requires that $|E_1|=4$ and $|E_2|=2$.  But the degree sequence of $X$ induced by $E_1$ is (2,2,0), in some order, and by $E_2$ is (1,1,0), in some order.  Together, no permutation of the coordinates of these two sequences can produce a sum of (2,2,2), which is the degree sequence of $X$ induced by all of $G$.  Hence, $r(K_{3,2})=3$.

To show that $r(K_{3,4})=4$, it suffices to show that $r(K_{3,4}) \nleq 3$.  Suppose, to the contrary that $E_1, E_2$ and $E_3$ form a regular partition of $K_{3,4}$.  If any $\langle E_i \rangle$ is isomorphic to $K_{3,3}$, say $\langle E_1 \rangle \cong K_{3,3}$, then $r(K_{3,4})=1+r(K_{3,1}) = 4$.  Hence, each $\langle E_i \rangle$ is either empty, 1-regular, or 2-regular.  So, for $i=1,2,3$, $|E_i| \in \{0,1,2,3,4,6\}$.  But $|E|=12$.  Hence, the partition of 12 edges into 3 parts can only take the forms 6+6, 6+4+2, 6+3+3, or 4+4+4.  When any $E_i$ has 6 edges, it contributes $(2,2,2,0)$ in some order to the degree sequence of $Y$.  However, two such sequences, say $(2,2,2,0)$ and $(2,0,2,2)$, do not add up $(3,3,3,3)$, the degree sequence of $Y$ in $G$. Hence, the form 6+6 is not possible.  Similar arguments show that the other three forms are also not possible.  Hence, $r(K_{3,4})=4$.  A similar counting argument as for $K_{3,4}$ shows that $r(K_{3,5})=4$.  This completes the basis step of the inductive proof.

Now fix $n \equiv 1, 2$~(mod~3).  Assume the assertion holds for smaller values of $n$.  Let $k = \left \lfloor n/3 \right \rfloor$ and let $s = r(K_{3,n})$.  To show that $s=k+3$, we now assume that $s \le k+2$ and arrive at a contradiction.  Let $E_1,\ldots,E_s$ be a minimum regular partition of $K_{3,n}$, where $s=k+2$ and $E_i$ can possibly be the empty set.  If any $\langle E_i \rangle$ is 3-regular, say $\langle E_1 \rangle$ is 3-regular, then $\langle E_1 \rangle \cong K_{3,3}$, and therefore, by the inductive hypothesis, $s = r(K_{3,n})$ $=1+r(K_{3,n-3})=1+(k-1+3)=k+3$, a contradiction. So now assume that each $\langle E_i \rangle$ is either empty, 1-regular, or 2-regular.  Then $|E_i| \in \{0,1,2,3,4,6\}$, where $|E_i| \in \{1,2,3\}$ if $\langle E_i \rangle$ is 1-regular. Let $\ell$ denote the number of $E_i$'s that have exactly 6 edges.  Each of the remaining $k+2-\ell$ $E_i$'s contain at most 4 edges.  Since these $k+2$ $E_i$'s together contain all $3n$ edges, it follows that $6 \ell + 4(k+2- \ell) \ge 3n$.  We now consider 2 cases, depending on whether $n \equiv 1$ or $n \equiv 2$ (mod~3).

Case 1: $n \equiv 1$ (mod~3). Suppose $n=3k+1$.  Then the above inequality on $\ell$ implies that $\ell \ge 2k-2 + \left \lceil \frac{k-1}{2} \right \rceil$. Each of the $\ell$ $E_i$'s that contain 6 edges contributes $(2,2,2,0,\ldots,0)$ to the degree sequence $(3,3,\ldots,3)$ of $Y$.  The three coordinates contributed by different $E_i$'s must be disjoint.  Hence, $3(2k-2+\left \lceil \frac{k-1}{2} \right \rceil) \le n$. This implies that $k=1$, a case that has been resolved already in the basis step.  Thus, for $k \ge 2$, we have that $s=k+3$.

Case 2: $n \equiv 2$ (mod~3).  Suppose $n=3k+2$.  Then the above inequality on $\ell$ implies that $\ell \ge 2k-1+\left \lceil k/2 \right \rceil$. As in Case 1, in order for the degree sequence conditions to not be violated, we require that $3(2k-1+\left \lceil k/2 \right \rceil) \le n$.  This implies that $k=0$.  Hence, for all $k \ge 1$, $s=k+3$ again.

\hfill\qed
\section{Some general bounds on the regular number}
\label{sec:bounds}

We now establish some simple on the regular number of a graph.  Let $\omega(G)$ denote the maximum size of a clique in $G$.  Then, since the edges of a maximum clique induce a regular subgraph of the graph, and since each of the remaining edges which are not in the clique induce a 1-regular subgraph, it follows that $r(G) \le |E| - {{\omega(G)} \choose 2} + 1$.  Similarly, let $\nu(G)$ denote the maximum size of a matching in $G$.  Then, $r(G) \le |E| - \nu(G) + 1$.  For each of these two bounds, there exist families of graphs which meet the bound with equality.

Before stating the next result, let us recall that the degree sequence of a graph is the sequence of non-negative integers listing the degrees of the vertices of $G$. For example, $K_{1,4}$ has degree sequence $(1,1,1,1,4)$, which contains two distinct elements: 1 and 4.

\begin{Proposition}
Let $\rho$ denote the number of distinct elements in the degree sequence of $G$.  Then $r(G) \ge \left \lceil \log_2 \rho \right \rceil$.
\end{Proposition}

\noindent \emph{Proof: }
Since every element $E_i$ in a regular partition of the graph induces a subgraph $\langle E_i \rangle$ whose vertices have the same degree $k_i$, the subgraph $G_i=(V,E_i)$ has a degree sequence containing at most two distinct elements: $k_i$ and possibly also 0. By adding $r$ such degree sequences, one corresponding to each subset $E_i$, we get a degree sequence that contains at most $2^r$ distinct elements.  Hence, the degree sequence of $G$ contains at most $2^{r(G)}$ distinct elements.
\hfill\qed.

\bigskip \noindent Let $\chi'(G)$ denote the chromatic index of $G$.  A fundamental result on the chromatic index of graphs, due to Vizing \cite{Vizing:1964}, states that $\chi'(G) \le \Delta+1$, where $\Delta$ denotes the maximum degree of a vertex in $G$.  Since $\chi'(G) \ge \Delta$ trivially, it follows that for every graph, $\chi'(G)$ equals either $\Delta$ or $\Delta+1$.  Graphs for which $\chi'(G)=\Delta$ are called Class 1 graphs, and graphs for which $\chi'(G)=\Delta+1$ are called Class 2 graphs.  Since $r(G) \le \chi'(G)$, Vizing's theorem immediately implies the following result.

\begin{Corollary}
For any graph $G$, $r(G) \le \Delta+1$.
\end{Corollary}

\bigskip \noindent If $G$ is disconnected, say, $G$ is the disjoint union of a 5-cycle and a path of length 3, then it is possible that the result holds with equality. However, preliminary calculations seem to indicate that, in fact, $r(G) \le \Delta$ might hold for all \emph{connected} graphs.

\bigskip \noindent We call the bound $r(G) \le \Delta$ (for connected graphs) the \emph{degree bound}.   To investigate if there exists a counterexample to the degree bound, we must examine Class 2 graphs, since the degree bounds holds for all Class 1 graphs.  Many of the known Class 2 graphs happen to be regular, and hence their regular number is trivially equal to 1. Furthermore, it can be seen that if $G$ contains a Hamilton cycle, then the degree bound holds.  To prove this result, observe that if $C$ is the edge set of a Hamilton cycle in $G$, then $\langle C \rangle $ is 2-regular and $\Delta(G-C) = \Delta(G)-2$.  Hence, $$r(G) \le 1 + r(G-C) \le 1+\chi'(G-C) \le 2+\Delta(G-C) = \Delta(G).$$  Thus, many of the known results on sufficient conditions for a graph to be Hamiltonian also ensure that the degree bound holds for those graphs. Even more, for this proof on the degree bound to go through, it is not necessary that $G$ contain a Hamilton cycle.  It suffices if $G$ contains a disjoint union of cycles that pass through all the vertices of $G$ that have degree $\Delta$ or $\Delta-1$.
We have shown that any connected graph that violates the degree bound is necessarily irregular, non-bipartite, non-Hamiltonian and Class 2.  Such graphs are extremely rare; in fact, almost all graphs are known to be Class 1 \cite{Erdos:Wilson:1977}.

\section{Concluding remarks}
\label{sec:conclusions}

A preliminary study of the regular number of some families of graphs was carried out, and some results and bounds were discussed. Many open questions remain.   It would be worthwhile to determine whether the degree bound $r(G) \le \Delta$ holds for all connected graphs $G$.  Furthermore, perhaps there is an upper bound, say $u(G)$, better than the degree bound, such that $r(G) \le u(G)$ for almost all graphs.  

Unlike in the case of the chromatic index, removing an edge can increase the regular number of a graph. For example, $r(K_n)=1$, and $r(K_n-e)$ equals 2 if n is even and 3 if n is odd.  An open problem is to investigate the maximum possible value of $r(G-e)$.

It is known that the problem of determining whether $\chi'(G)$ equals $\Delta$ or $\Delta+1$ is NP-complete \cite{Holyer:1981}.  The problem of determining the complexity of the regular number for various families of graphs is open.  Another interesting problem is to design algorithms to decompose a given (bipartite) graph into the minimum number of regular subgraphs.

We have stated above our conjectures related to the regular number of complete bipartite graphs and our conjecture on the degree bound.

\bibliographystyle{plain}
\bibliography{refsregnum}

\end{document}